\documentclass[useAMS,usenatbib]{mn2e}
\usepackage{graphicx}
\usepackage{longtable}
\newcommand{\Pdot}{{\ifmmode{\dot{P}}\else{$\dot{P}$}\fi}}
\newcommand{\mns}{\!\!\!}

\title[Binaries with High Period Change Rates]{ASAS Eclipsing Binaries with Observed High Period Change Rates}
\author[B. Pilecki, D. Fabrycky, R. Poleski]
{B. Pilecki$^{1}$\thanks{e-mail:  pilecki@astrouw.edu.pl},
D. Fabrycky$^{2}$\thanks{e-mail:  dfab@astro.princeton.edu},
R. Poleski$^{1}$\thanks{e-mail:  rpoleski@astrouw.edu.pl}\\
$^{1}$Warsaw University Observatory, Al. Ujazdowskie 4, PL-00-478, Poland \\
$^{2}$Princeton University Observatory, Peyton Hall, New Jersey 08544, USA}

\date{Accepted --.
      Received -- ;
      in original form --}

\pubyear{2007}

\begin{document}

\maketitle

\label{firstpage}

\begin{abstract}
We present 31 bright eclipsing contact and semi-detached binaries that showed high period change rates in a 5 year interval in observations by the All-Sky Automated Survey (ASAS).  The time-scales $|P / \Pdot|$ of these changes range from only 50 up to 400 kyr. The orbital periods of 10 binaries are increasing and of 21 are decreasing, and even a larger excess is seen in contact binaries, where the numbers are 5 and 17, respectively.
Period change has previously been noticed for only two of these binaries; our observations confirmed a secular period drift for SV Cen and period oscillations for VY Cet. The spectroscopic quadruple system V1084 Sco shows both period change and brightness modulation.
All investigated binaries were selected from a sample of 1711 (1135 contact and 576 semi-detached) that fulfilled all criteria of data quality.
We also introduce a ``branch'' test to check if luminosity changes on part of the binary's photosphere has led to a spurious or poorly characterized period change detection.

\vspace{2cm}
\end{abstract}


%%%%%%%%%%%%%%%%%%%%%%%%%%%%%%%%%%%%%%%%%%%%%%%%%%
\section{Introduction}  
\label{sect:intro}
%%%%%%%%%%%%%%%%%%%%%%%%%%%%%%%%%%%%%%%%%%%%%%%%%%

There are various reasons for an observed period change of an eclipsing binary star.
\begin{itemize}
\item[-] During conservative mass exchange, the period decreases if the mass loser is currently more massive than the gainer and increases if the opposite is true.
\item[-] Mass lost from the system in a rather isotropic wind causes the period to increase as the specific angular momentum increases.
\item[-] Mass lost in a flattened configuration may carry away angular momentum, causing the period to decrease.
\item[-] Stellar spin angular momentum may be transfered to the orbit by tidal dissipation, causing the period to either increase or decrease until a spin equilibrium is reached in which the tidal torque vanishes.
\item[-] Period oscillations are also possible, caused by oblateness changes during magnetic cycles as proposed by Applegate (1992).
\end{itemize}

However, apparent orbital period change is not always an intrinsic phenomenon and sometimes is not even physical.  Virtual oscillations of period are observed due to the light-time effect (LITE) caused by the displacement induced by one or more additional companions (Irwin 1959, Mayer 1990, Pribulla et al. 2005).  Also, for a minority of stars, surface activity resulting in apparent phase shift of parts of the light curve may be misinterpreted as a period change. This phase shift is, however, generally smaller than 0.02 of a period and does not accumulate (Kalimeris, Rovithis-Livanou \& Rovithis 2002).

The Algol paradox, in which the less massive component of a binary is more evolved, seems to be resolved (see Pustylnik 1998 for a review), but the structure and evolution of close interacting binaries still hide some mysteries (Ibanoglu et al. 2006, Webbink 2003).  The origin of contact binaries is still being debated (Yakut \& Eggleton 2005, Eker et al. 2006, Pribulla \& Rucinski 2006, St\c{e}pie\'n 2006, Van Hamme 2006).  It is also unclear whether close binaries evolve through angular momentum loss (AML) (van't Veer 1979, Rahunen 1981, Vilhu 1982, St\c{e}pie\'n 1995 and others), which requires mass loss in magnetized winds, or thermal relaxation oscillations (TRO) (Flannery 1976, Lucy 1976, Robertson \& Eggleton 1977, Wang 1999, Webbink 2003), which requires mass exchange between components, or a mixture of both (Qian 2003). TRO predicts that contact binaries can even become semi-detached for periods of time.

A large sample of stars with swift period changes can therefore shed light on the evolution of interacting binaries, give information about stellar structure and magnetic fields involved in the Applegate mechanism, or signal the presence of a multiple system in LITE.

Recently Paczy{\'n}ski et al. (2006) (hereafter Paper I) has brought attention to a great number of eclipsing binaries from the ASAS Catalogue (Pojma\'nski, Pilecki and Szczygiel 2005). The variable stars were discovered quasi-uniformly for declination
${\rm < + 28^{\circ} }$, covering almost 3/4 of the full sky.
For brightness within $8 < V < 12$ mag nearly all the variable stars with the amplitude $\Delta V > 0.1$ mag have been found (see Paper I for details).  There are 5,384 eclipsing contact binaries (EC), 2,957 semi-detached binaries (ESD) and 2,758 detached binaries (ED) in the catalogue.

The precision of a period change rate measurement of a sinusoid is:
\begin{equation}
\sigma_{\Pdot} \approx 3 \frac{P^2}{N^{1/2} T^2} \frac{\sigma}{A}, \label{eqn:sigpdot}
\end{equation}
where $P$ is the period and $A$ is the full-amplitude of the sinusoid, $N$ is the number of data points (roughly uniformly spread over a duration of $T$), and $\sigma$ is the photometric precision of each data point.  This estimate comes from a Fisher-matrix analysis in which the second derivative matrix of the model with respect to its parameters is inverted to find the errors on those parameters.  Typical numbers for ASAS light curves are $P=0.4$ d, $A=0.4$ mag, $N=400$, $T=1900$ d, $\sigma=0.02$ mag.  Therefore for a typical curve an optimistic $3\sigma$ detection of period change is $\sim 1.5 \times 10^{-6}$ d/yr, comparable to period changes discovered through O-C diagrams spanning many decades obtained by literature searches (e.g. Qian 2001a,2001b).

In our examinations we consider only EC and ESD binaries (i.e., we neglect ED).  We require more than 300 observation points, randomly distributed in time, and a binary period shorter than 10 days (see section 2 for details). Section 3 describes the methods used to find period changes. The 31 binaries with high period change rates are presented in section 4. For discussion of results see section 5.

%----------------------- Fig. 1 ---------------------------------------
\begin{figure}
\includegraphics[width=\linewidth]{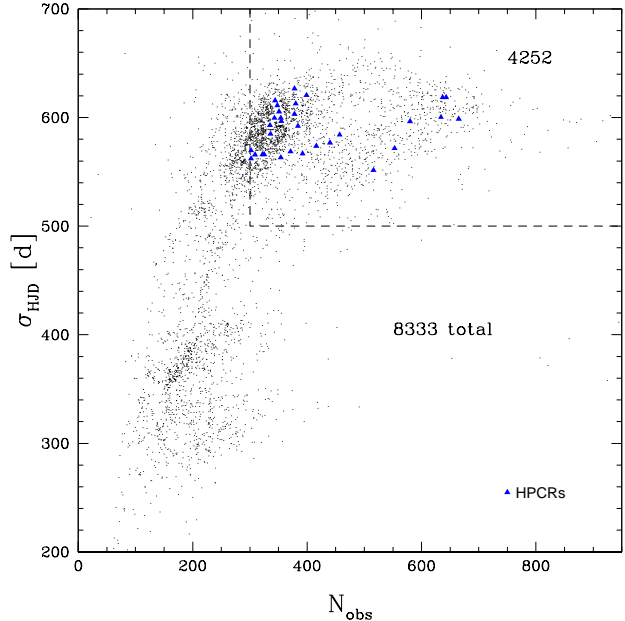}
\caption{
This diagram shows the number of points ($N_{obs}$) and time-scatter of observations ($\sigma_{HJD}$) for all 8333 EC and ESD binaries.  Low $\sigma_{HJD}$ means that observation points are too highly concentrated. Stars selected for analysis (4252) are enclosed with a dashed line. Triangles mark detected high period change rate (HPCR) stars.
}\label{fig:nsig}
\end{figure}
%----------------------- Fig. 1 ---------------------------------------

%%%%%%%%%%%%%%%%%%%%%%%%%%%%%%%%%%%%%%%%%%%%%%%%%%
\section{Data Selection}
\label{sect:data}
%%%%%%%%%%%%%%%%%%%%%%%%%%%%%%%%%%%%%%%%%%%%%%%%%%

In this section we will focus only on the selection criteria. For detailed information on data statistics and quality please refer to Paper I.

Good coverage of the phase with observations is essential to get reliable results, therefore, we have selected only stars that have no fewer than \mbox{$N_{obs} = 300$} observations with acceptable quality flags (A or B).  Moreover, while examining light curves of ASAS eclipsing binaries we found that some of the stars (even with a large number of points) have most observations concentrated around specific HJD (Heliocentric Julian Date).  This situation is not good for a period change study, as the $T^{-2}$ factor in equation \ref{eqn:sigpdot} attests.  Therefore, we required the observation time standard deviation,
\begin{equation}
\sigma_{HJD} = \sqrt{ \frac{1}{N_{obs}} \sum_i (t_{HJD,i} - \overline{t}_{HJD})^2 }, \label{eqn:sighjd}
\end{equation}
where $\overline{t}_{HJD}$ is the average of all observation times, be greater than $500$~d, which corresponds to about 1750 days of uniform observations.  A larger value of $\sigma_{HJD}$ gives a better constraint on period changes as observation times are less concentrated. All binaries in the parameter space ($N_{obs}$,$\sigma_{HJD}$) are presented in Fig. \ref{fig:nsig}.  We analyzed the binaries enclosed with a dashed line. These selection criteria reduced the number of stars from 8333 to 4252.

An upper period limit was set to 10 days to cover at least 200 orbital cycles, while the maximum is about 10000 cycles. The quality of period change determination, however, progressively diminishes with increasing period (equation \ref{eqn:sigpdot}). This is illustrated by Table \ref{tab:list}, where we present stars for which the highest accuracy period changes were obtained: the longest period there is only $1.658$~d. There are fewer long period systems, so this criterion reduced the number of binaries only by 272 to 3980 systems (2573 contact, 1407 semidetached).
To make sure that the period change analysis would give reliable results this set was further reduced to 1711 (1135 EC, 576 ESD) stars with highest signal to noise ratio. This step is described in the next section.

%%%%%%%%%%%%%%%%%%%%%%%%%%%%%%%%%%%%%%%%%%%%%%%%%%
\section{Method description}
\label{sect:method}
%%%%%%%%%%%%%%%%%%%%%%%%%%%%%%%%%%%%%%%%%%%%%%%%%%
We applied two methods to determine the period change and a third method to check whether these changes are real. They are described in following subsections.

\subsection{Period change} \label{sect:periodchange}
\vspace{9 pt}
{\it Local Scatter Reduction}

In the first method we attempt to minimize brightness scatter along the light curve to make it as smooth as possible.  Before making calculations a linear trend in brightness was removed from the data.  A two parameter search was performed with power defined as {\small $POW(P,\Pdot) = \sigma_g / \sigma_l(P,\Pdot)$}, where $\sigma_g$ is global standard deviation and $\sigma_l$ is local standard deviation of all brightness measurements.  The global standard deviation is calculated about the average value of all brightness measurements, $\overline{m}$.  We calculate $\sigma_l$ for assumed values for $P$ and $\Pdot$ by the following algorithm. For each point $m_i$ $(i=1,...,N)$ we choose nearby points that are closer than $\epsilon=5/N$ in phase. We then calculate the average value of these points, called $\overline{m_i}$, which is used to calculate the standard deviation $\sigma_l$ of the quantity $(m_i - \overline{m_i})$. Therefore we have:
\begin{equation}
POW(P,\Pdot) = \sqrt{\frac{\sum_i (m_i - \overline{m})^2}{\sum_i (m_i - \overline{m_i}(P,\Pdot))^2} }.
\end{equation}

For each star we then chose a model of $P$ and $\Pdot$ with the highest power, which is the best fit. Stars with power less than 2.5 were discarded because the signal to noise ratio is too low. This step greatly reduced the number of stars from 3980 to 1711 but was necessary to avoid many spurious detections.

%----------------------- Fig. 2 ---------------------------------------
\begin{figure}
\begin{tabular}{c}
\includegraphics[width=\linewidth]{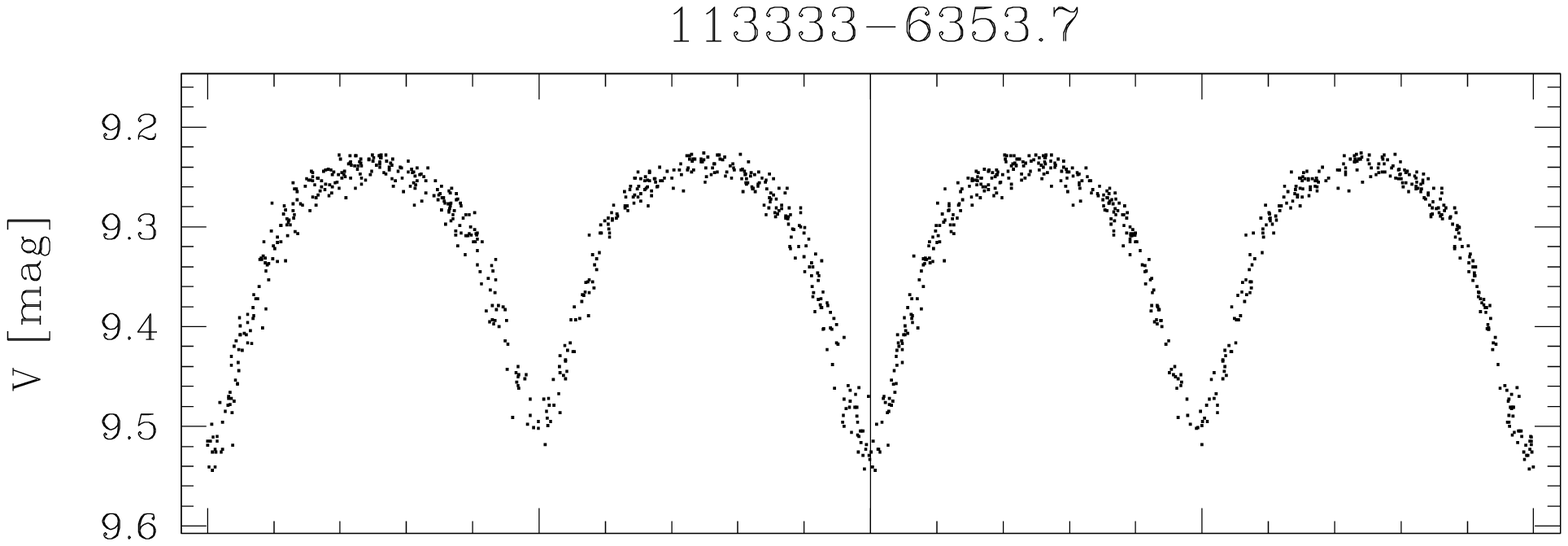}\\
\includegraphics[width=\linewidth]{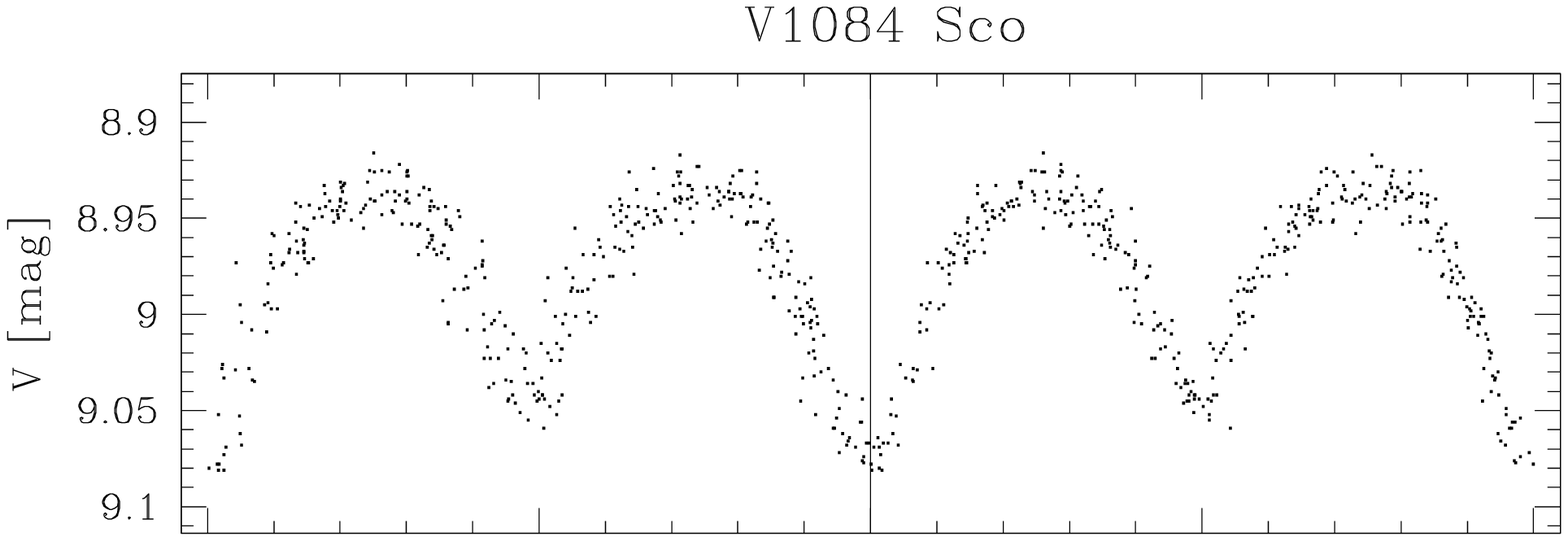}\\
\end{tabular}
\caption{
Two examples of a light curve phased with best $P$ achieved in one-parameter model (left) and best $P$ and $\Pdot$ from two-parameter model (right).  The biggest improvements are seen at the steepest parts of the light curves.
}\label{fig:metex}
\end{figure}
%----------------------- Fig. 2 ---------------------------------------

In the same way as above, we have found the best one-parameter model for which $\Pdot$ was set to zero and we searched for the best period $P$. Next, powers of these two models were compared and improvement defined as {\small $IMP=POW_{max}(P,\Pdot)/POW_{max}(P,0)$} was evaluated. If it was higher than a cutoff level the star was selected as a high period change rate (HPCR) object. This level was set according to results from a synthetic light curve analysis, to ensure that in fewer than $0.5\%$ of trials a model with $\Pdot=0$ will have a detected period change.  As the examined sample consists of 1711 stars, we are aware that up to about 8 stars may be spurious detections.

Finally we were left with 31 HPCR stars.  The light curves of ASAS 113333-6353.7 and V1084 Sco phased with and without taking the period change into account are presented in Figure \ref{fig:metex}, as examples.

\vspace{9 pt}
\noindent {\it Harmonic fits}

Using the model:
\begin{eqnarray}
\label{eqn:harmonic} 
X(t) & = & \mu + \alpha t + \beta(t^2 - T^2/12) \\
     &   & - \sum_{m=0}^H A_m \cos\left( m \theta(m, t) \right), \nonumber
\end{eqnarray}
where
\begin{equation}
\theta(m, t) = \omega (t-T_{0,m}) + \dot{\omega} (t^2/2 - T^2/24), \label{eqn:phase}
\end{equation}
we fit ($2H+5$) parameters ($[A_m, T_{0,m}]_{m \in [1,H]}, \mu, \alpha, \beta, \omega, \dot{\omega}$) by least-squares.  $H$ is the number of harmonics, which was chosen as six.  $T$ is the range of observations times for the particular light curve, and $t$ are the observation times relative to the center of that observed interval.  Note that $\theta$ only depends on the order of the harmonic $m$ through the phase $T_{0,m}$ of that sinusoid.  The form of this model was determined to be ideal for minimizing covariance between the parameters.  First, we solve for the binary's angular frequency (mean motion) $\omega$ on a finely-spaced grid by fitting models with $H$ harmonics that are linear in the parameters (the coefficients of sines and cosines).  At this stage the mean magnitude $\mu$ is fitted, but $\alpha$, $\beta$, and $\dot{\omega}$, which are the linear and quadratic drifts in magnitude and the linear drift of $\omega$, respectively, are fixed at zero.  

This procedure gives values to initialize the Levenberg-Marquardt algorithm (Press et al. 1989), which is a non-linear least-squares solver.  The initial epochs $T_{0,m}$ are chosen near the center of the dataset, consistent with the sine and cosine coefficients, whereas $\alpha$, $\beta$, and $\dot{\omega}$ are initialized at zero.  The data points are weighted by the inverse of the square of their estimated photometric errors.  All the parameters are allowed to vary as the algorithm minimizes the sum of the squares of the residuals.

The algorithm converges in $\sim6-10$ iterations to return the best-fitting values and formal errors for the parameters.  The reduced chi-squared value of the residuals indicates whether or not the model fits well.  For some light curves, it is considerably different from 1, which means either that the estimated photometric errors are not good, or that the model does not explain all the variation intrinsic to the light curve.  Also, we multiply the parameters' formal errors by the square root of the reduced chi-squared, which produces more realistic errors and is equivalent to rescaling the photometry's errors so that the reduced chi-squared equals 1.

\begin{figure}
\includegraphics[width=\linewidth]{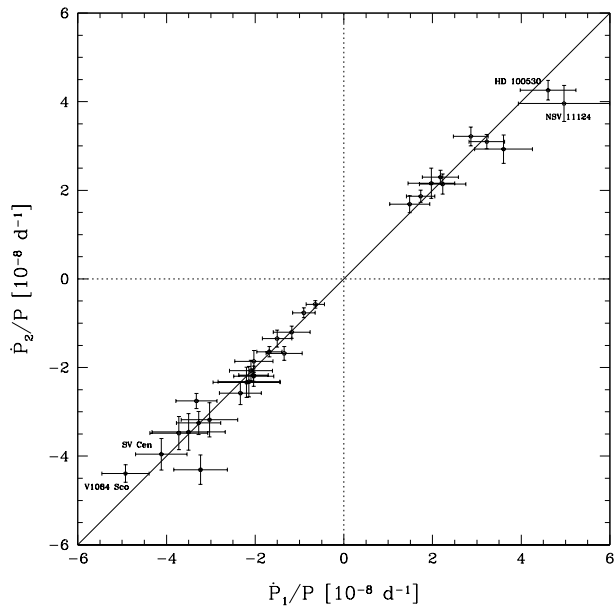}
\caption{
Comparison of inverted time-scales obtained with the two different methods described in Section \ref{sect:periodchange}. $\Pdot_1$ was found with the \emph{Local Scatter Reduction} method, and $\Pdot_2$ with the \emph{Harmonic fits} method.  Stars with decreasing period have negative values while ones with increasing period have positive values. For each case two stars with the shortest time-scales are labeled. It is convincing that both quite independent methods give similar results.
}\label{fig:ts-ts}
\end{figure}

We performed this second method on the HPCR objects selected by the first method. Results of both methods are compared in Fig. \ref{fig:ts-ts}. Error bars are $1\sigma$ errors determined from a synthetic light curve analysis for the first method and the rescaled formal errors for the second one.

%----------------------- Fig.3 ---------------------------------------
%----------------------- Fig. 3 -------------------------------------

\subsection{Branch Test}  \label{sec:bt}

A standard graphical tool for evaluating whether or not the period is constant is the observed minus calculated (O-C) method (see Zhou 1999 for a review).  The observed times of minima, referenced to calculated time of minima based on assuming a constant period, are plotted as a function of time.  In such diagrams, a constant period results in a straight line; if the correct period is assumed, this line has zero slope.  However, a period that is drifting at a linear rate will produce a parabola on an O-C diagram.  Residuals from straight lines in O-C diagrams are sometimes attributable to flux variations instead of genuine period changes, perhaps due to star spots generated by the strong magnetic fields in close binaries (see Kalimeris, Rovithis-Livanou \& Rovithis 2002 for a good analysis of this phenomenon).  Nevertheless, flux fluctuations are not likely correlated over the whole photosphere of the binary, so they are not supposed to cause identical apparent O-C variations at all phases of the binary light curve.

\begin{figure}
\includegraphics[width=\linewidth]{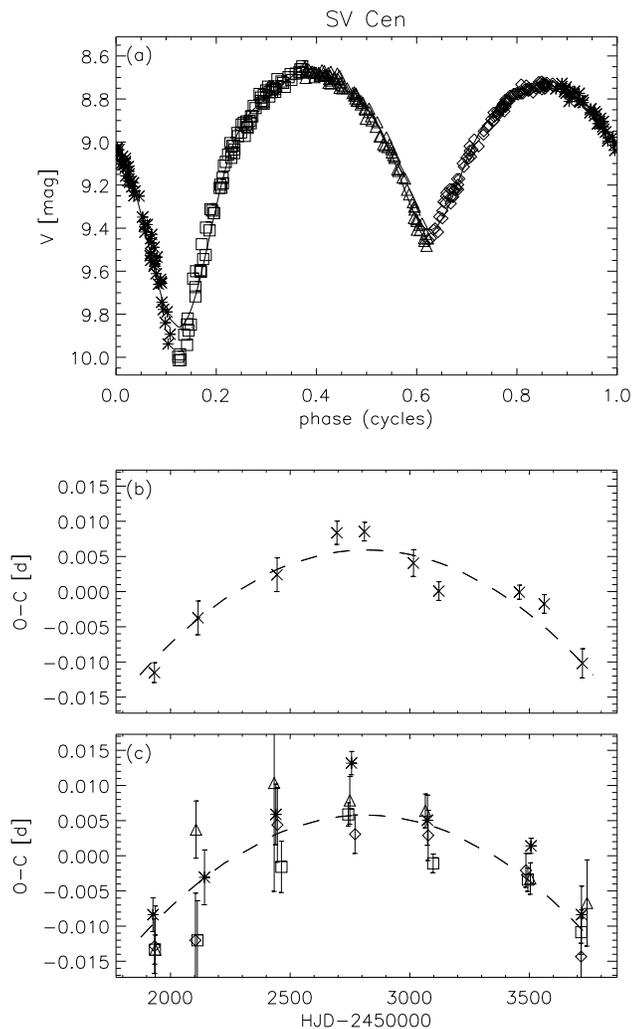}
\caption{
A binary, SV Cen, that passes the branch test.  (a) The phased light curve with points assigned to one of four branches (marked by different symbols) and a line indicating the best-fitting six harmonic ($H=6$) model.  (b) An O-C diagram (see section \ref{sec:bt} regarding its construction) for all the data.  Error bars are determined by phase shifting the model away from the best fitting solution until $\Delta \chi^2 = 1$.  The dashed line indicates the best-fitting linearly-changing period model.  (c) An O-C diagram for each of the branches individually (see section \ref{sec:bt}).  
}\label{fig:btex1}
\end{figure}

\begin{figure}
\includegraphics[width=\linewidth]{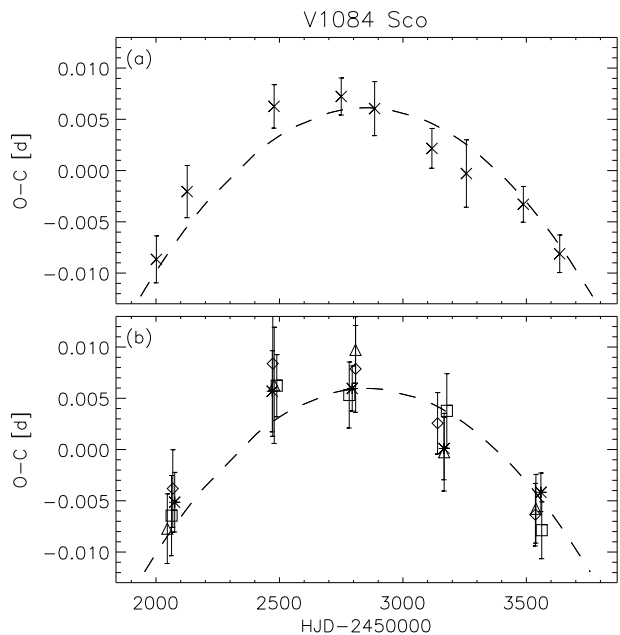}
\caption{
Panels a) and b) are the same as b) and c) in Fig. \ref{fig:btex1}; a phased light curve is presented in Fig. \ref{fig:lc_tab}.  This binary, V1084 Sco, also passes the branch test, and its O-C diagram deviates less from the best-fitting model than that of SV Cen. It means that the period change rate is relatively constant in time.
}\label{fig:btex2}
\end{figure}

\begin{figure}
\includegraphics[width=\linewidth]{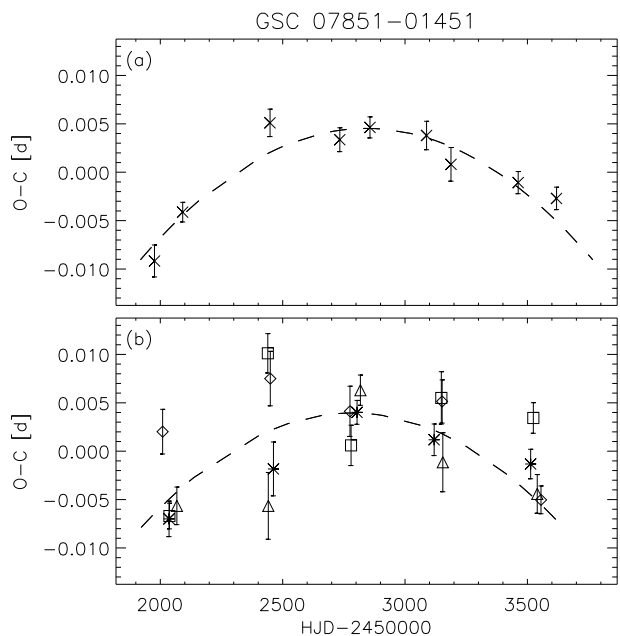}
\caption{
Panels a) and b) are the same as b) and c) in Fig. \ref{fig:btex1}; a phased light curve is presented in Fig. \ref{fig:lc_tab}.  Despite having a clean-looking O-C diagram when all the data are considered, this binary, GSC 07851-01451, fails the branch test because the O-C for each branch follows a different pattern.  Scattered data or surface activity may explain these features.
}\label{fig:btex3}
\end{figure}

We developed a test to determine if flux variations on part of the binary's photosphere could be causing what we have interpreted as period changes, or at least perturbing their values.  First, the light curve is fitted by a six-harmonic model (equations \ref{eqn:harmonic} and \ref{eqn:phase}) with the periods listed in Table \ref{tab:list}.  A parabolic brightness change was removed from the light curve, such that $\alpha$ and $\beta$ in equation \ref{eqn:harmonic} were forced to zero, but we checked that this had little effect on the following results.  The photometric errors were rescaled as described in the previous section regarding the harmonic fits method.  Starting with the phase of the primary minimum, four boundaries are set on the folded light curve, spaced by a quarter cycle.  These boundaries establish four ``branches'' in phase for which we individually plot and compare O-C diagrams; therefore we call this method the branch test.  These branches shall be referred to by $b=1,2,3,4$, with $b=1$ including the brightening after primary minimum and the others in chronological order (see Fig. \ref{fig:btex1}a for an example).  Bins are also established along the time axis, which in chronological order shall be labeled $j=1,2,...,n$.  Local values of O-C as a function of $b$ and $j$ are determined using groups of $\geqslant 10$ points in each branch within each time bin, giving a total of $4n$ O-C values for the $4n$ groups defined from the total $N_{obs}$ photometric data points.  The best-fitting six-harmonic model is shifted in phase to achieve least-squares residuals for each group, giving a value for O-C.  An O-C error, $\sigma_{O-C}$, is determined by shifting the phase away from that best fit until the chi-squared value for the model fitting the data of the group increases by one from the minimum value.  Figures \ref{fig:btex1}--\ref{fig:btex3} show the branch test applied to three stars in our sample.

%----------------------- Fig. 4 ---------------------------------------
%----------------------- Fig. 4 ---------------------------------------

%----------------------- Fig. 5 ---------------------------------------
%----------------------- Fig. 5 ---------------------------------------

%----------------------- Fig. 6 ---------------------------------------
%----------------------- Fig. 6 ---------------------------------------

We identified a statistic to quantify if the light curve shape remains constant while its phase is shifting.  That is, in section \ref{sect:periodchange} we found stars for which the function \mbox{$(O-C)(b,j)$} has a roughly quadratic dependence on $j$; now we wish to know if it has a statistically significant dependence on branch number $b$.  Thus this analysis is similar to a two-way analysis of variance with interaction (Ross 2000).

The mean O-C value for each time bin was computed using the following formula:

\begin{equation}
\overline{(O-C)}_{b} (j) = \frac{ \sum_{b=1}^{4} (O-C)(b,j) \sigma_{O-C}^{-2}(b,j) }{ \sum_{b=1}^{4} \sigma_{O-C}^{-2}(b,j)} . \label{eqn:averageomc}
\end{equation}

Then we may compute:

\begin{equation}
\chi^2 = \sum_{b=1}^{4} \sum_{j=1}^{n} \left[(O-C)(b,j) - \overline{(O-C)}_{b} (j)\right]^2 \sigma_{O-C}^{-2}(b,j), 
\end{equation}
which will be distributed as a $\chi^2$ random variable with $3n$ degrees of freedom if all the branches agree, that is if 
$$(O-C)(b,j) = \overline{(O-C)}_{b} (j)$$
for all $b$ and $j$.  A total of $n$ degrees of freedom are used in empirically estimating the values of $\overline{(O-C)}_{b} (j)$ through equation \ref{eqn:averageomc}.  If the branches do not agree, then the computed value of $\chi^2$ will lie on the extreme tail of the distribution, with only a fraction $p_{BT}$ of the distribution at more extreme values.  Therefore, if the hypothesis that the branches agree is true, the probability of drawing such an extreme result randomly is approximately $p_{BT}$.  We reject that hypothesis if $p_{BT} < 0.01$,
which means that light curve shape changes (e.g. local photometric variations due to star spots) probably perturb true phase shifts or even imitate them in extreme cases.

We recognize that this branch test is not entirely statistically rigorous, but we find that the value $p_{BT}$ gives a quantitative measure consistent with the subjective visual examination of each O-C diagram.

%%%%%%%%%%%%%%%%%%%%%%%%%%%%%%%%%%%%%%%%%%%%%%%%%%
\section{Results}
\label{sect:results}
%%%%%%%%%%%%%%%%%%%%%%%%%%%%%%%%%%%%%%%%%%%%%%%%%%

Here we present the results of the methods of section 3, supplemented by some additional measurements and data published by others.

From all the examined binaries the local scatter reduction method identified 31 stars with significant period changes, which are listed in Table \ref{tab:list}.  Due to the relatively short time baseline of ASAS observations and limited accuracy, all these stars have either a high or very high period change rate ($\mathbf{\Pdot}$ column). Estimated $1\sigma$ errors for these rates are quoted in parentheses (which is the error in the last one or two digits). A small `b' there means that a star failed the branch test.

Of the stars that failed the branch test, three of them (ASAS 062254-7502.0, ASAS 002449-2744.3, and GSC 07851-01451---whose branch test is shown in Fig. 6) have phase shifts that span more than $\sim4\%$ of a period, so according to Kalimeris, Rovithis-Livanou \& Rovithis (2002) the phase shifts are probably not induced by spots.  Perhaps in these cases there is surface activity that merely adds systematic error to our measurement of the period change rates.  In the other cases labeled `b', the change is small enough, and our time baseline is short enough, that it may be possible for surface activity to cause the apparent period change.

The fastest detected period change ($\Pdot=-2.5\times10^{-5}$ d/yr) was found for SV Cen, whose period change is well known, but its time-scale defined as $|P/\Pdot|$ is longer than for three other stars because it has also the longest period in the sample. V1084 Sco and NSV 11124 have the shortest time-scales of the sample, about $55$ kyr each.  The slowest reported period change is $+6.5\times10^{-7}$ d/yr for NSV 4657, which also has the longest time-scale of over $400$ kyr.

It is interesting that amongst HPCR (High Period Change Rate) objects there are two times more negative period changes than positive ones (21 compared to 10), a ratio that we discuss in the final section.

Six objects from our selection have a known tertiary component (according to the Washington Double Star Catalog; Mason et al. 2001), but because these are visual doubles they cannot cause the measured period changes by the light-time effect or by dynamical effects.  However, these stars are not resolved by ASAS and the third light affects the system brightness and its amplitude -- those cases are marked by `$\Downarrow$' in the \textbf{Amp} column (note that similar effects must be considered while evaluating the LITE hypothesis for any star).  One spectroscopic quadruple is marked in the same way (see Section 4.3 for details). Also, nine stars were identified as X-ray sources in ROSAT data (Voges et al. 1999, Voges et al. 2000), which may imply increased activity in the system. Both cases are marked in \textbf{St} column of Table \ref{tab:list}.

%======================= Table 1 ======================================
\begin{table*}
\caption{We list here 31 HPCR stars with their time of first minimum after HJD=2451868 ($HJD_0$), orbital period (P), period change rate ($\Pdot$; error in the last one or two digits is quoted in parentheses, `b' marks failed branch test), corresponding time-scale ($|P/\Pdot|$), V band maximum brightness ($V_{max}$; followed by `{\tiny$\uparrow$}' if rising, `{\tiny$\downarrow$}' if descending) and amplitude (Amp - a primary minimum depth; with `{\tiny$\Downarrow$}' if third light is present), minima depth ratio ($d_S/d_P$), O'Connell effect ($\Delta max$; if greater than $2\%$ of Amp), spectral type (Sp), status (St; `X' means this is also a ROSAT X-ray source while `3' or `4' denotes number of known components if more than 2, `s' marks spectroscopic multiple) and cross identification (GCVS name if available, other if not; was known as variable if bolded). First column is ASAS identification (was not known as variable if bolded). }
\label{tab:list}
\begin{tabular}{|c|c|c|r@{.}l|c|c|c|c|c|c|c|c|}
\hline
ID & $HJD_0$ & P & \multicolumn{2}{c}{$\dot{P}$} & $|P/\dot{P}|$ & $V_{max}$ & Amp & $d_S/d_P$ & $\Delta max$ & Sp & St & Other ID\\
 & [d] & [d] & \multicolumn{2}{c}{\tiny [$10^{-5} \frac{d}{yr}$]} & [kyr] & [mag] & [mag] &  & [mmag] &  &  & \\
\hline
114757-6034.0 & 1.030 & 1.657589 & -2&50{\scriptsize(35)} & 66 &  8.67 & 1.30 & 0.58 & 65 & B2   & --   & \bfseries SV Cen \\
\sffamily\bfseries 113333-6353.7 & 0.478 & 0.991121 & 1&67{\scriptsize(23)$^b$} & 59 &  9.23 & 0.29{\tiny$\Downarrow\mns$} & 0.90 & -- & B8   & 4    & HD 100530 \\
184110-7229.7 & 0.391 & 0.710832 & 1&29{\scriptsize(27)} & 55 & 11.60 & 0.73 & 0.36 & -- & --   & --   & \bfseries NSV 11124 \\
071225-2530.0 & 0.346 & 0.720825 & 0&85{\scriptsize(11)$^b$} & 85 &  9.41 & 0.40 & 0.73 & -- & --   & --   & \bfseries VW CMa \\
\hline
\sffamily\bfseries 074537-3109.6 & 0.398 & 0.602926 & -0&77{\scriptsize(19)} & 78 & 10.73{\tiny$\downarrow\mns$} & 0.27 & 0.80 & 41 & --   & --   & GSC 07106-00494 \\
\sffamily\bfseries 144910-4424.3 & 0.191 & 0.446193 & -0&61{\scriptsize(11)} & 74 & 10.37{\tiny$\downarrow\mns$} & 0.19{\tiny$\Downarrow\mns$} & 0.86 & -11 & --   & 3    & DON 684 \\
231524-5018.4 & 0.152 & 0.418344 & 0&55{\scriptsize(10)} & 76 & 11.48 & 0.43 & 0.99 & -- & --   & --   & \bfseries NSV 14467 \\
173758-3911.4 & 0.293 & 0.303315 & -0&54{\scriptsize(5)} & 56 &  8.93 & 0.14{\tiny$\Downarrow\mns$} & 0.79 & -- & G6V  & X4$_s$ & \bfseries V1084 Sco \\
\hline
\sffamily\bfseries 062426-2044.9 & 0.231 & 0.384692 & -0&47{\scriptsize(7)} & 82 & 10.64{\tiny$\uparrow\mns$} & 0.34 & 0.96 & -- & --   & --   & GSC 05959-01748 \\
\sffamily\bfseries 102014-1351.6 & 0.285 & 0.381025 & -0&45{\scriptsize(9)} & 85 & 10.28 & 0.12 & 0.97 & -11 & --   & --   & BD-13 3091 \\
\sffamily\bfseries 082456-4833.6 & 0.180 & 0.364875 & -0&44{\scriptsize(7)} & 84 & 11.63 & 0.34 & 0.96 & -- & --   & --   & -- --- \\
\sffamily\bfseries 160302-3749.4 & 0.354 & 0.363239 & -0&40{\scriptsize(9)$^b$} & 90 & 10.92{\tiny$\uparrow\mns$} & 0.38 & 0.91 & 27 & --   & X    & GSC 07851-01451 \\
\hline
\sffamily\bfseries 004717-1941.6 & 0.031 & 0.488810 & -0&39{\scriptsize(13)} & 130 & 11.21 & 0.38 & 0.69 & -- & --   & --   & CPD-20 88 \\
\sffamily\bfseries 060557-5342.9 & 0.137 & 0.463634 & 0&38{\scriptsize(9)} & 120 & 10.82 & 0.32 & 0.89 & -- & --   & --   & GSC 08521-01468 \\
\sffamily\bfseries 231603-1553.5 & 0.143 & 0.470110 & -0&37{\scriptsize(12)$^b$} & 130 &  9.98{\tiny$\uparrow\mns$} & 0.27 & 0.93 & -11 & G0   & --   & HD 219462 \\
\sffamily\bfseries 070959-3639.5 & 0.165 & 0.371829 & -0&28{\scriptsize(7)} & 130 &  9.66{\tiny$\downarrow\mns$} & 0.18{\tiny$\Downarrow\mns$} & 0.92 & -- & F5V  & X3   & HD 55100 \\
\hline
\sffamily\bfseries 065232-2533.5 & 0.247 & 0.418639 & 0&27{\scriptsize(5)$^b$} & 160 &  8.61{\tiny$\downarrow\mns$} & 0.38 & 0.87 & 15 & F6V  & X    & HD 50494 \\
\sffamily\bfseries 004430-3606.5 & 0.064 & 0.246537 & 0&26{\scriptsize(4)} & 96 &  9.58 & 0.13{\tiny$\Downarrow\mns$} & 0.80 & -11 & G5   & X3   & HD 4227 \\
014933-1937.6 & 0.145 & 0.340809 & -0&25{\scriptsize(4)} & 130 & 11.00 & 0.64 & 0.93 & -- & G5V  & --   & \bfseries VY Cet \\
135243-5532.5 & 0.112 & 0.580784 & -0&25{\scriptsize(9)} & 230 &  9.44 & 0.53 & 0.57 & 27 & B9IV & --   & \bfseries V758 Cen \\
\hline
\sffamily\bfseries 072729-5056.5 & 0.317 & 0.330557 & 0&24{\scriptsize(6)} & 140 & 11.89 & 0.37 & 0.97 & 18 & --   & X    & -- --- \\
\sffamily\bfseries 002449-2744.3 & 0.265 & 0.313661 & -0&23{\scriptsize(5)$^b$} & 140 & 12.39 & 0.76 & 0.82 & -- & --   & --   & -- --- \\
\sffamily\bfseries 002821-2904.1 & 0.195 & 0.269892 & -0&23{\scriptsize(5)} & 120 & 11.97 & 0.54 & 0.80 & -37 & --   & --   & -- --- \\
\sffamily\bfseries 062254-7502.0 & 0.148 & 0.257707 & 0&21{\scriptsize(4)$^b$} & 130 & 11.40 & 0.43 & 0.92 & -- & --   & --   & -- --- \\
\hline
\sffamily\bfseries 025016-4649.2 & 0.139 & 0.271753 & -0&20{\scriptsize(4)} & 140 & 12.44 & 0.51 & 0.95 & -12 & --   & --   & -- --- \\
\sffamily\bfseries 093312-8028.5 & 0.118 & 0.406067 & -0&20{\scriptsize(6)} & 200 & 10.64 & 0.33 & 0.94 & -- & --   & --   & GSC 09404-00233 \\
\sffamily\bfseries 144047-3725.3 & 0.276 & 0.353410 & 0&19{\scriptsize(6)} & 180 &  9.25 & 0.28{\tiny$\Downarrow\mns$} & 0.93 & 8 & G2V  & X3   & HD 128910 \\
195350-5003.5 & 0.110 & 0.286827 & -0&18{\scriptsize(3)$^b$} & 160 & 11.25 & 0.99 & 0.69 & -- & --   & X    & \bfseries NSV 12502 \\
\hline
\sffamily\bfseries 052851-3010.2 & 0.153 & 0.302101 & -0&17{\scriptsize(4)} & 180 & 11.30 & 0.36 & 0.96 & -- & --   & --   & -- --- \\
071727-4007.7 & 0.144 & 0.320265 & -0&11{\scriptsize(3)} & 300 & 11.14 & 0.66 & 0.85 & -- & --   & X    & \bfseries GZ Pup \\
095048-6723.3 & 0.235 & 0.276943 & -0&07{\scriptsize(2)$^b$} & 430 & 11.15 & 0.79 & 0.92 & -17 & --   & --   & \bfseries NSV 4657 \\
\hline
\end{tabular}

\end{table*}
%======================================================================

\begin{figure}
\includegraphics[width=\linewidth]{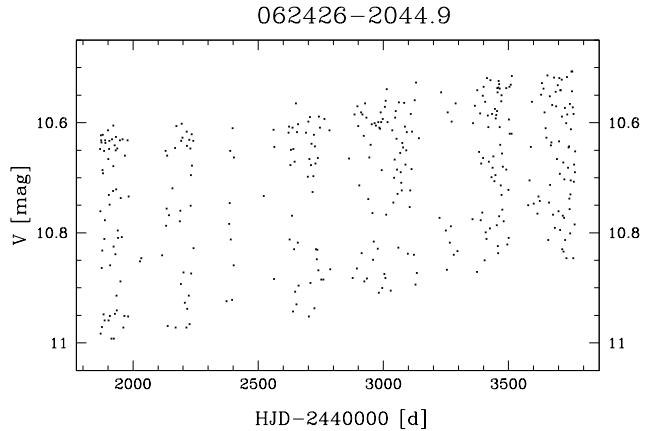}
\caption{
The raw light curve of 062426-2044.9, which has the most prominent trend in average brightness that is roughly equal to $0.025$ mag/yr. This star has also a quite large negative period change.
}\label{fig:rawlc}
\end{figure}

During data analysis for seven stars a significant trend in average brightness was removed. In Table \ref{tab:list}, $\mathbf{ V_{max}}$ for these stars is marked with `$\uparrow$'
(if rising) or `$\downarrow$' (if decreasing).  The extreme case of 062426-2044.9 with a linear trend of $+0.025$ mag/yr (over $7\%$ of the amplitude per year) is presented in Fig. \ref{fig:rawlc}.

Because the Applegate mechanism predicts luminosity variation with the same period as the orbital period modulation, it is possible that the linear brightness and period changes have a common mechanism. If this is true, we will have a unique means to verify this rarely confirmed effect (as for CG Cyg by Afsar et al. 2004 and WW Cyg by Zavala et al. 2004) for a substantial number of stars.  But we have only detected linear changes in period and luminosity, whereas the Applegate mechanism is oscillatory in nature.  Continued monitoring of these binaries is necessary, which ASAS intends to do.

%----------------------- Fig. 7 ---------------------------------------
%----------------------- Fig. 7 ---------------------------------------

\begin{figure}
\includegraphics[width=\linewidth]{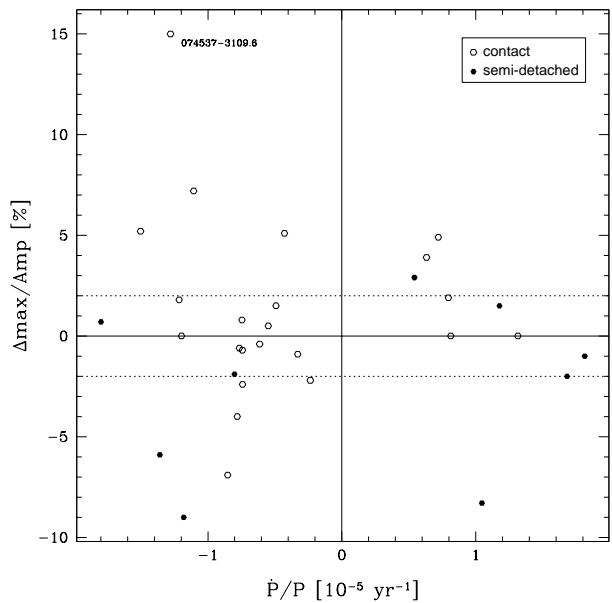}
\caption{
Magnitude of the O'Connell effect (normalized to the amplitude) versus period change rate (normalized to the period). Dotted lines represent significance level used to quote the effect in Table \ref{tab:list}. Overall statistics did not agree with expected ratio of positive to negative effect of 3:1 and small asymmetry is seen between stars with different sign of period change.  It also appears that in the high period change rate regime, contact binaries preferentially have decreasing periods, and this should be confirmed with larger statistics.
}\label{fig:ts_ocon}
\end{figure}

\begin{figure}
\begin{tabular}{l}
\includegraphics[width=\linewidth]{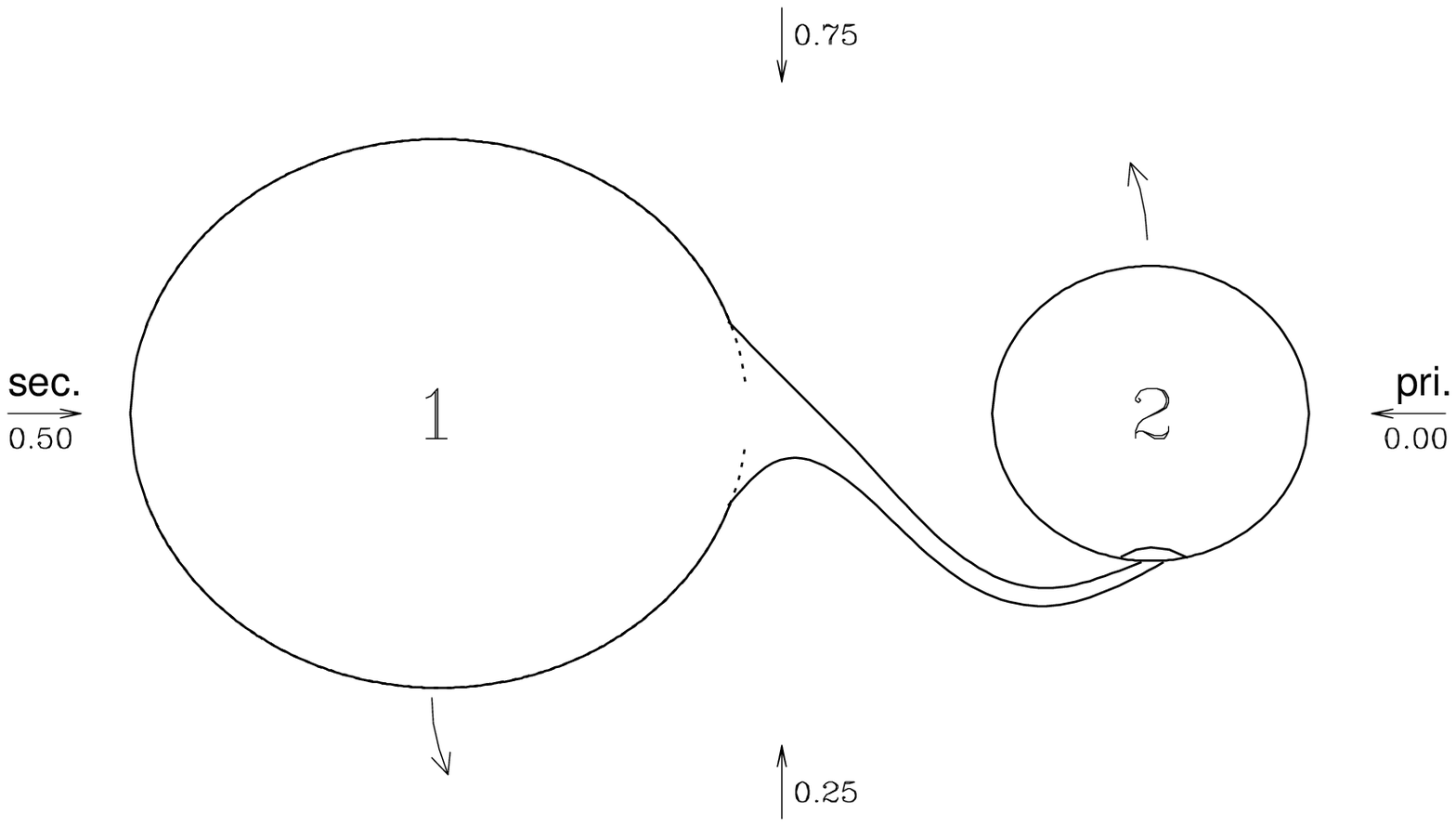}\\
\includegraphics[width=\linewidth]{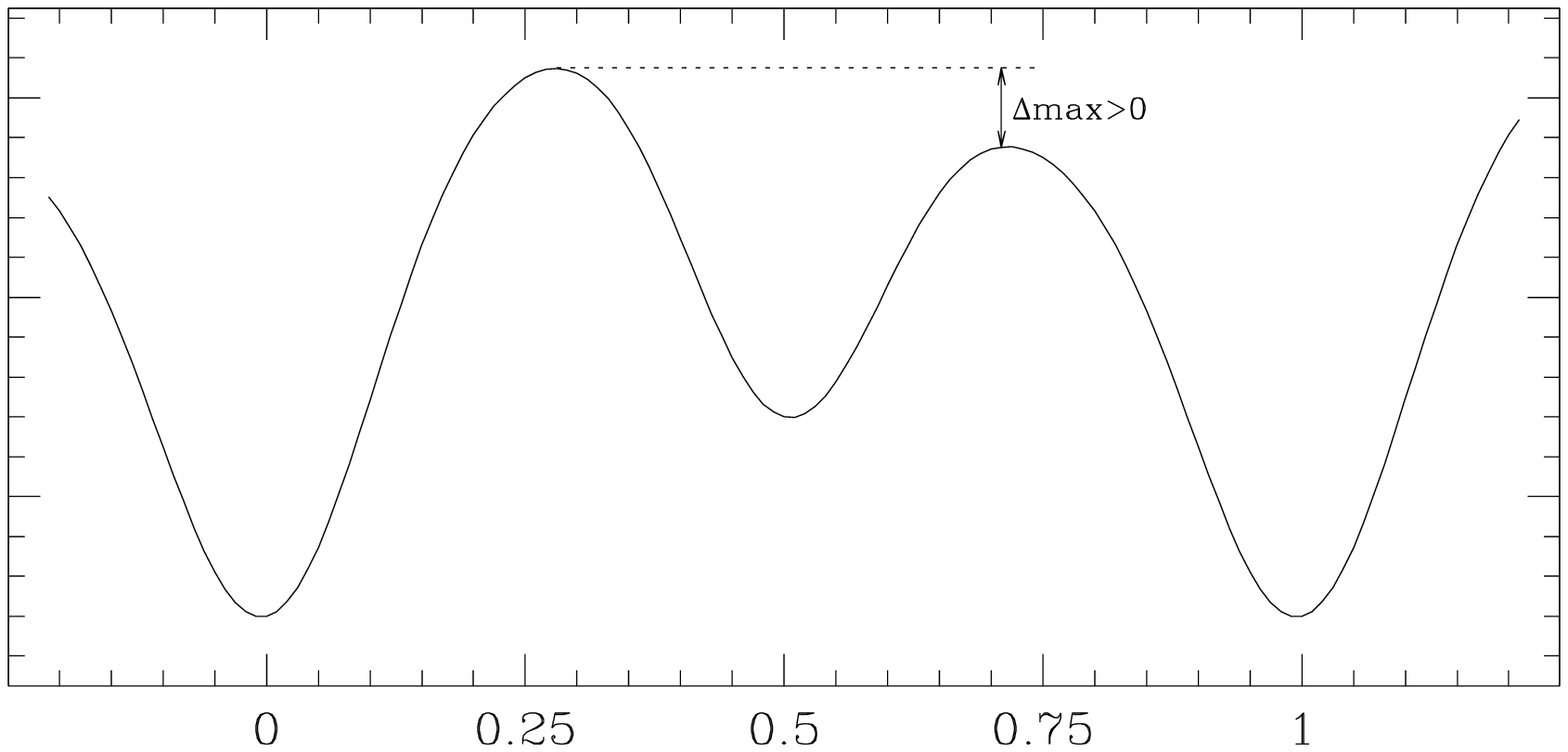}\\
\end{tabular}
\caption{
An example of positive O'Connell effect.  (Top) A stream of matter flowing from the hotter component (1) and heating a surface of its companion (2).  As the binary rotates, four viewing directions and their phases are marked with arrows.  (Bottom) The lightcurve that arises from such a situation, with a positive $\Delta max$, illustrates how mass transfer is a potential mechanism for this phenomenon.  If the component (1) is the cooler one, a negative effect can be observed.  Another explanation of the O'Connell effect are dark spots, e.g. if component (2) has a dark spot in the same place as the hot spot in the picture above, the light curve will have a negative effect.
}\label{fig:ocon_ex}
\end{figure}

The O'Connell (1951) effect (see also Davidge \& Milone 1984), in which the two maxima have different brightness, was also analyzed and quoted in the table if significant ($\mathbf{\Delta max} > 0.02 \times Amp$), and it may be important according to Fig. \ref{fig:ts_ocon}.
Let us define the positive effect as the one in which a maximum after a primary minimum is higher than a maximum after a secondary minimum and the negative effect in which the opposite is true (see Fig. \ref{fig:ocon_ex} for a possible mechanism of the former).  Using data presented in Paper I we have checked what the global statistics for our selected 1711 binaries are.  Table \ref{tab:ocon_stat} contains the numbers of positive and negative effects for these stars as well as their ratios.

%----------------------- Fig. 8 ---------------------------------------
%----------------------- Fig. 8 ---------------------------------------

%----------------------- Fig. 9 ---------------------------------------
%----------------------- Fig. 9 ---------------------------------------

%======================= Table 2 ======================================
\begin{table}
\caption{The statistics of the O'Connell effect for 1711 stars selected for our analysis are presented for the whole sample (EC+ESD) and for each type of eclipsing binaries separately. Positive effect is marked by `+' and negative by `$-$'. The numbers and ratios of the binaries showing these effects are given when the effect is significant ($\Delta max > 0.02 \times Amp$) and (in parentheses) regardless of its magnitude (ie. for all 1711 stars). } \label{tab:ocon_stat}
\centering
\begin{tabular}{cccc}
\hline $\Delta max$ & EC & ESD & EC+ESD \\
\hline
 + & 378 (770) & 89 (341) & 467 (1111) \\
 $-$ & 100 (365) & 66 (245) & 166 (600) \\
 ratio +:$-$& 3.78 (2.11) & 1.35 (1.4) & \textbf{2.87 (1.85)} \\
\hline
\end{tabular}
\end{table}
%======================================================================

One can see that the ratio for stars of both types is almost 3:1 when only significant effects are taken into account and almost 2:1 for all effects.
Amongst our HPCR objects there are, however, 7 stars with positive and 7 with negative effect, suggesting rather a 1:1 ratio. Nevertheless the sample has only a slight statistical significance even if we include stars below a $2\%$ limit seen on the diagram, for which the same 1:1 ratio applies. The measurements of the O'Connell effect in the column $\mathbf{\Delta max}$ of Table \ref{tab:list} are positive for a positive effect and negative for a negative one.

The ASAS classification (see Pojma\'nski 2002 for details on classification method) was redone for the HPCR stars, including new data and a period change.  If multiple types were reported only the most probable one was used. There are 22 contact and 9 semi-detached binaries.  

This preliminary classification is based solely on {\it V}-band photometry, and is therefore uncertain. The GCVS classification (Samus, 2004) is, however, available only for six of these stars and is of no use for statistics. Both classifications are given in Table \ref{tab:vartype}.

When using the ASAS classification, roughly the same number of ESD stars have increasing and decreasing period (5 and 4, respectively), but among the EC binaries there are 17 stars with decreasing and only 5 with increasing period, as seen in Fig. \ref{fig:ts_ocon}. The classification is not biased by whether the period is increasing or decreasing, so this large difference may be shown to be real if confirmed in a larger sample.  Note that we explore a high period change rate domain only and the statistics for the whole population may be different.

%======================= Table 3 ======================================
\begin{table}
\caption{
Variability type of HPCR objects. Renewed ASAS classification and GCVS type is given.}\label{tab:vartype}
\begin{tabular}{cr@{,}l c cr@{,}l}
\hline
ID & \sc asas & \sc gcvs  &\hspace{-1em} $|$&	ID & \sc asas & \sc gcvs  \\
\hline
114757-6034.0 & EC  & EB &\hspace{-1em} $|$& 065232-2533.5 & EC & -- \\
113333-6353.7 & ESD & -- &\hspace{-1em} $|$& 004430-3606.5 & ESD & -- \\
184110-7229.7 & ESD & -- &\hspace{-1em} $|$& 014933-1937.6 & EC & EW \\
071225-2530.0 & ESD & EB &\hspace{-1em} $|$& 135243-5532.5 & EC & EW \\
\hline
074537-3109.6 & EC & -- &\hspace{-1em} $|$& 072729-5056.5 & EC & -- \\
144910-4424.3 & ESD & -- &\hspace{-1em} $|$& 002449-2744.3 & EC & -- \\
231524-5018.4 & EC & -- &\hspace{-1em} $|$& 002821-2904.1 & EC & -- \\
173758-3911.4 & ESD & EW &\hspace{-1em} $|$& 062254-7502.0 & EC & -- \\
\hline
062426-2044.9 & EC & -- &\hspace{-1em} $|$& 025016-4649.2 & EC & -- \\
102014-1351.6 & ESD & -- &\hspace{-1em} $|$& 093312-8028.5 & EC & -- \\
082456-4833.6 & EC & -- &\hspace{-1em} $|$& 144047-3725.3 & ESD & -- \\
160302-3749.4 & EC & -- &\hspace{-1em} $|$& 195350-5003.5 & EC & -- \\
\hline
004717-1941.6 & ESD & -- &\hspace{-1em} $|$& 052851-3010.2 & EC & -- \\
060557-5342.9 & EC & -- &\hspace{-1em} $|$& 071727-4007.7 & EC & EW\\
231603-1553.5 & EC & -- &\hspace{-1em} $|$& 095048-6723.3 & EC & -- \\
070959-3639.5 & EC & -- &\hspace{-1em} $|$&  &  & \\
\hline
\end{tabular}

\end{table}
%======================================================================

Ratio of stars with positive and negative O'Connell effect calculated separately for EC binaries (6:4) is still far from the value for the whole sample (4:1). There are too few ESD binaries to make a similar statement for them, but we noticed that they seem to avoid large positive effect, especially while decreasing period.  It is also interesting that in our sample there is no EC star with both increasing period and negative O'Connell effect.

One may doubt if period changes detected during the relatively short observations carried out by ASAS are real. In this set there are, however, two stars already known as period changing binaries with suitably high period change rates. They represent two kinds of period change. SV Cen has an exceptionally strong secular period change, while the period of VY Cet oscillates with a $\sim7.3$ year cycle.  These examples show what is detectable by ASAS, but other possibilities (e.g. non-periodic oscillations or abrupt change) are possible.  Also, recent spectroscopic observations suggest that V1084 Sco is a multiple system, so we may suspect LITE as the period change mechanism. We describe these three stars in more detail in the following subsections.

\subsection{SV Cen}
\label{ssect:SVCen}

This example shows that with the very short time baseline of ASAS (when compared to the usual O-C diagram time span of decades) we can still detect secular changes of period if the effect is strong. SV Cen has been studied thoroughly (Paczy\'nski 1971, Rucinski 1976, Herczeg \& Drechsel 1985, Rucinski et al. 1992 and many others), and its period variability is well known and clearly seen on an O-C diagram (Kreiner, Kim \& Nha 2001). 

\begin{figure}
\includegraphics[width=\linewidth]{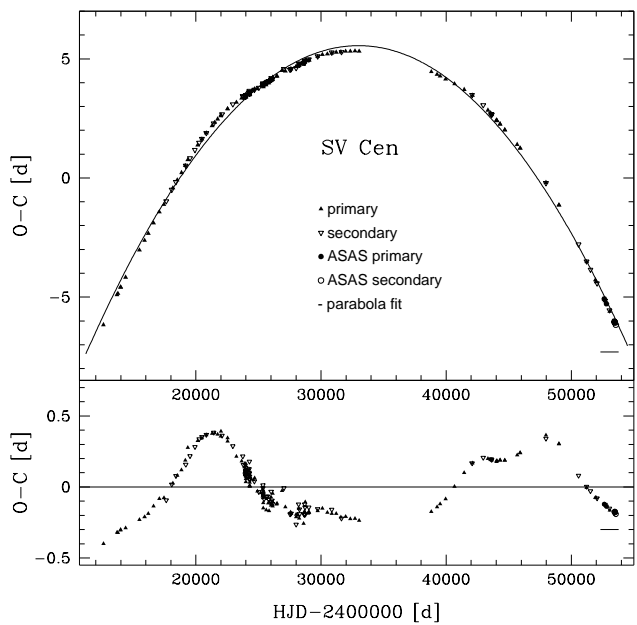}
\caption{
A known O-C diagram (triangles) for star SV Cen extended with 7 new minima times from ASAS data (circles) calculated with linear (upper) and parabolic (lower) ephemeris. A strong secular trend is clearly seen on the upper panel with small irregular oscillations superimposed on it, which are better seen on the lower panel. New minima times (above a short horizontal line) fit well to older data.
}\label{fig:SVCen.O-C}
\end{figure}

In Fig. \ref{fig:SVCen.O-C} we present an O-C diagram (Kreiner, private comm.) with 7 new minima times from ASAS data (see Table \ref{tab:SVCen.mt}). The upper panel diagram was calculated with a linear ephemeris for the times of primary ($I$) minima:
$$ Min\, I = 2423794.5 + 1.6595 E $$
and the lower one with a parabolic ephemeris:
$$ Min\, I = 2423797.7473 + 1.660332 E -7.502\times10^{-8} E^2, $$
where $E$ is the cycle number.  New minima times fit well to older data.

A strong secular trend is clearly seen (upper panel) with small irregular oscillations (better seen on lower panel) superimposed on it. Drechsel et al. (1982) proposed that mass flowed from the less massive star.  Matter expelled from the outer Lagrangian point acts to decrease the orbital period, and matter falling onto the more massive companion acts to increase it.  The former dominates, but irregularities originate from a varying ratio of lost to transferred matter. Another proposal is that mass transfer occurs in the opposite direction and variations are just fluctuations in the transfer rate (Wilson \& Starr 1976).

%======================= Table 4 ======================================
\begin{table}
\caption{
Minima times for SV Cen from ASAS data.} \label{tab:SVCen.mt}
\centering
\begin{tabular}{cc}
\hline
HJD-2400000 & Min. \\
\hline
52634.8504 & pri \\
52810.5532 & pri \\
53433.8003 & pri \\
53467.7806 & sec \\
53482.7066 & sec \\
53511.7096 & pri \\
53560.5986 & sec \\
\hline
\end{tabular}
\end{table}
%======================================================================

The secular period change derived from the parabolic ephemeris is $-3.3\times 10^{-5}$ d/yr, but because of irregularity, the instantaneous value may differ considerably.  In particular, our value from Table \ref{tab:list} is $-2.5\times 10^{-5}$ d/yr, which is in agreement with the upward curvature of the last $\sim 7$ years of the quadratic ephemeris O-C curve (the lower panel of Fig. \ref{fig:SVCen.O-C}). A light curve of this star is in Figs. \ref{fig:btex1} and \ref{fig:lc_tab}.

Only high rates of period change due to rapid mass flows as in SV Cen are detectable with our method. As mass transfer is faster in stars with shorter thermal time-scales, we can only detect period changes in stars of early spectral types. In Table \ref{tab:list} we can see that three out of ten stars are as early as type B, so we expect this kind of period variability to be a minority in our sample.

%----------------------- Fig. 10 ---------------------------------------
%----------------------- Fig. 10 ---------------------------------------

\subsection{VY Cet} \label{ssect:VYCet}

The next example shows we can also detect period oscillation. VY Cet is a WUMa (GCVS) type contact binary whose period instability is clearly seen on an O-C diagram (Kreiner, Kim \& Nha 2001). The periodicity of its period change was studied by Qian (2003), who proposed it is due to LITE and found a period of the third component of around 7.3 years, with minimum mass $0.62 M_\odot$.  Another possibility is the Applegate mechanism, but it is less probable as VY Cet has a stable light curve and its period oscillations seem to be regular.

%----------------------- Fig. 11 ---------------------------------------
\begin{figure*}
\includegraphics[width=\linewidth]{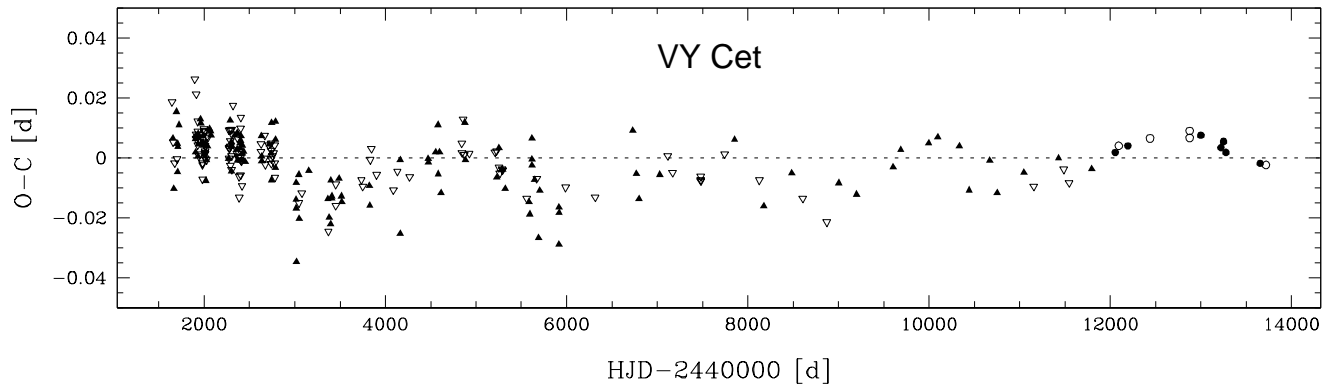}
\caption{
A known O-C diagram (triangles) for star VY Cet extended with new 12 minimum times (circles) from ASAS data. In the diagram oscillations are clearly seen superimposed on small secular trend and new minima fit well to older data. Older primary and secondary minima designations were swapped to agree with ASAS data.
}\label{fig:VYCet.O-C}
\end{figure*}
%----------------------- Fig. 11 ---------------------------------------

In Fig. \ref{fig:VYCet.O-C} we present an O-C diagram with 12 new minima times from ASAS (see Table \ref{tab:VYCet.mt}) which fit very well to older data (Kreiner, private comm.). Other points in the range covered by ASAS are discarded because of much larger scatter.  Older primary and secondary minima designations (which are ambiguous because the depth ratio is close to 1) were swapped to agree with the ASAS light curve.  The diagram was calculated with the following ephemeris for times of primary minima:

$$ Min\, I = 2441645.219 + 0.3408097 E. $$

The oscillations are clearly seen, and given the observational scatter and potential secondary period change mechanisms, their amplitude and phase stability is exemplary.
This observation cautions that many of the reported HPCR stars may actually have period oscillations with a time-scale comparable or longer than the time baseline of ASAS.  This interpretation may be favored relative to mass transfer for stars with late spectral type, whose thermal time-scale is significantly longer than the measured period change time-scale.

A light curve of this star is in Fig. \ref{fig:lc_tab}.

%======================= Table 5 ======================================
\begin{table}
\caption{
Minima times for VY Cet from ASAS data.} \label{tab:VYCet.mt}
\centering
\begin{tabular}{cc}
\hline
HJD-2400000 & Min. \\
\hline
52055.9347 & pri \\
52093.9372 & sec \\
52195.6688 & pri \\
52440.8840 & sec \\
52877.8022 & sec \\
52877.8044 & sec \\
53000.6649 & pri \\
53223.8912 & pri \\
53251.8396 & pri \\
53277.7375 & pri \\
53655.6918 & pri \\
53720.6155 & sec \\
\hline
\end{tabular}
\end{table}
%======================================================================

\subsection{V1084 Sco} \label{ssect:V1084Sco}

After we had selected V1084 Sco as a HPCR binary, we discovered that Rucinski \& Duerbeck (2006) reported that this star is probably a quadruple system. Using spectroscopic data obtained at the European Southern Observatory, they concluded that it consists of a detached binary accompanied by a contact binary responsible for photometric variation (due to the shape of its light curve, ASAS type is ESD). They also stated that this star would be very interesting if the mutual period of revolution of the two binaries is short enough to observe.

\begin{figure}
\includegraphics[width=\linewidth]{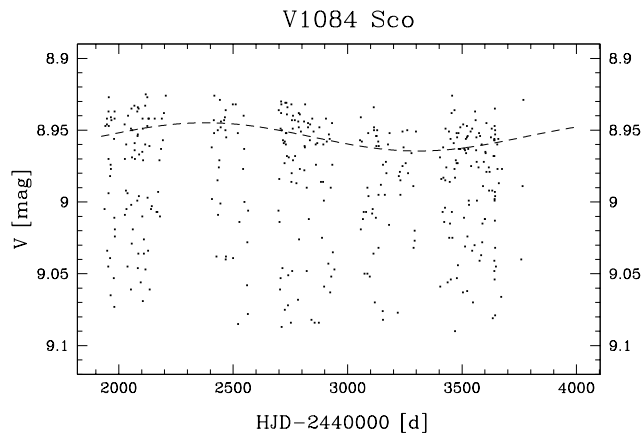}
\caption{
The raw light curve of V1084 Sco star with a sinusoid fitted to the maxima points. The brightness is modulated with about 5 year time-scale and an amplitude of 15 mmag.
}\label{fig:V1084Sco.raw}
\end{figure}

It is most likely that the period change that we have detected for V1084 Sco is due to LITE. Assuming the mass of the detached binary is less than $5 M_{\odot}$ we expect the mutual period to be less than 20 years.
There is also a brightness modulation seen for this star with a time-scale of about 5 years (see Fig. \ref{fig:V1084Sco.raw}).  This modulation cautions that an effect other than LITE may make an important contribution to the period change.

\begin{figure*}
\includegraphics[width=\linewidth]{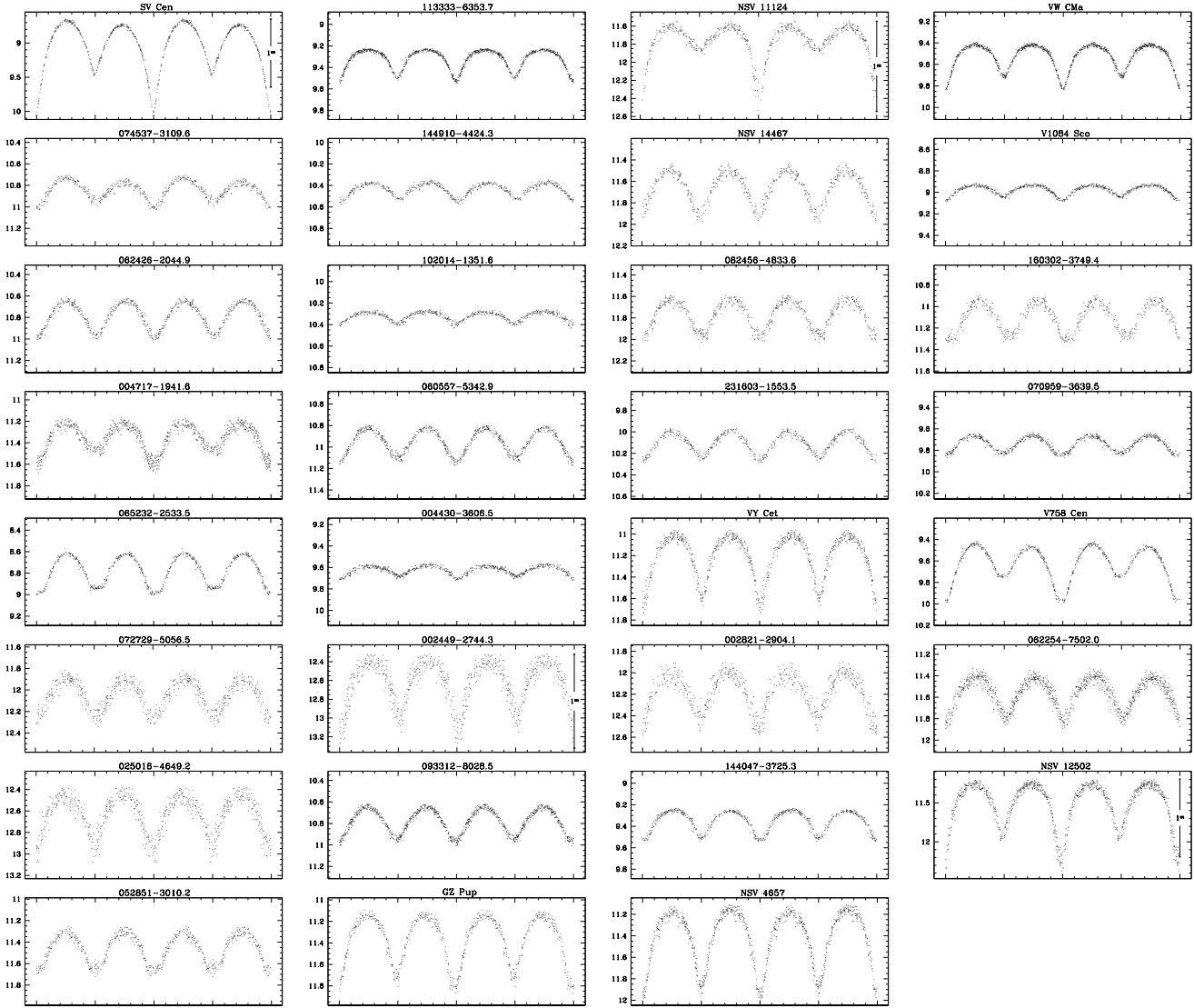}
\caption{
Composed of two cycles, these 31 light curves of HPCR stars are phased with the best model parameters of $P$ and $\Pdot$. Minimum magnitude span is $1$~mag, but in cases where it is higher, a $1 mag$-bar is plotted on the right side of each light curve. GCVS identification is given when possible, otherwise ASAS ID is used. A linear trend in brightness was removed where significant, by shifting data points to the magnitude level at the beginning of observations.  The light curves appear, left-to-right and top-to-bottom, in the same order as their corresponding stars appear in Table \ref{tab:list}.
}\label{fig:lc_tab}
\end{figure*}

A phased light curve of V1084 Sco is presented in Figs. \ref{fig:metex} and \ref{fig:lc_tab}. The O-C diagram is given in Fig. \ref{fig:btex2}.

%----------------------- Fig. 12 ---------------------------------------
%----------------------- Fig. 12 ---------------------------------------

%%%%%%%%%%%%%%%%%%%%%%%%%%%%%%%%%%%%%%%%%%%%%%%%%%
\section{Conclusions}
\label{sect:concl}
%%%%%%%%%%%%%%%%%%%%%%%%%%%%%%%%%%%%%%%%%%%%%%%%%%

Out of 8333 EC and ESD binaries from the ASAS Catalogue, 1711 have characteristics appropriate for a period change search. In this paper we have presented 31 binaries (22 EC and 9 ESD) with a high period change rate that is statistically significant given the photometric precision.  Because of the short time base of just over 5 years it is impossible to say whether these trends are secular.  The true time-scales are also uncertain: the rate of change may be the maximum of a generally slower process or an abrupt change may be averaged out to a lower value.  One of the shortest detected time-scales of about 55,000 years is for V1084 Sco - a quadruple system consisting of a pair of binaries. The highest period change rate is for SV Cen, for which $\Pdot=-2.5\times 10^{-5}$ d/yr.

The reported HPCR objects are $2\%$ of 1711 examined binaries.  Because ASAS is an ongoing project, every year of observations will increase the number of stars that will pass the selection algorithm and greatly improve the significance of the statistical trends.  With the scaling in equation \ref{eqn:sigpdot}, we may expect two times better precision for period change rate with about two more years of data, which will probably double the overall number of HPCR detections.  As a longer baseline and more data are collected, the number of binaries with data suitable for a period change search (here 1711) will also increase.

A 21:10 ratio of negative to positive period changes was found.  For binaries with statistically symmetrical period changes (like in LITE) it is obviously expected to be 1:1. The probability of obtaining a ratio at least this different from 1:1 can be computed from the binomial distribution; it is $7.1\%$, which is not quite statistically significant.  The difference can be attributed to other non-symmetric oscillations or secular changes.

One mechanism for period decrease at the observed rate is mass transfer from an early type star on a thermal time-scale.  If the percentage of early type stars (Table \ref{tab:list}) whose spectral types are known from the literature is extrapolated to the whole sample, then about ten stars have the required thermal time-scale.  The rest of twenty late-type stars should then have a roughly symmetrical distribution of period change as a portion of cyclic phenomena, and this mixture of mechanisms might explain the 2:1 ratio. The problem is that one of the three known early type stars in the sample has an increasing period, but it can be overcome if the real proportion of spectral types is different than that assumed above or if the real ratio of period change directions is closer to 1:1.

The observed asymmetry is more striking if we look at contact and semi-detached systems separately. ESD stars are evenly distributed between positive and negative period change values, and so the ratio for EC stars rises to 17:5. Assuming real ratio is 1:1, ratios that extreme have a probability of only $1.7\%$, which is marginally statistically significant.  The explanation from the previous paragraph cannot be applied here so simply. What we need now is a non-symmetrical mechanism for contact binaries of mostly late type that is stronger than simple mass transfer on thermal time-scales.  A truly secular mechanism is not needed; it is sufficient to have a cyclic phenomenon in which the period decreases more quickly than it increases and is thus detected only in the former stage.

It is also worth noting that a recent study of period change phenomenon for the contact binaries from the OGLE database (Kubiak et al. 2006) suggests a 1:1 ratio of significant detections for these systems. However, their much longer timespan (up to 14 years) allowed them to securely detect more stars with lower period changes rates than those found here. If we set a threshold at $\Pdot=2 \times 10^{-6}$, above which about $85\%$ of our detections are placed, we can see a similar excess of the negative changes (8 compared to 3) in their results. Besides, eight of our stars have higher rates than the maximum detected in that analysis ($5 \times 10^{-6}$).  This comparison encourages the conclusion that the period change distribution is truly non-symmetrical in the high period change rate regime. The addition of their 11 HPCR objects to our 31 makes the probability of 29:13 (or more extreme) ratio lower than $2.0\%$.

Mass transfer is a possible cause of the O'Connell effect, but it also causes period changes, so it is not surprising that some interesting correlations are seen among high period change rate stars.  This result is not yet statistically significant, but we hope to check these trends with more data.  In systems where LITE is the cause of period changes, virtually no change to the O'Connell effect (relative to the general population) is anticipated.

For some stars we discovered luminosity changes accompanying period variations. This may be important for confirming the Applegate magnetic cycles as the cause of variations but needs at least a few more years of observations. These discoveries could be interesting as the latest studies of this mechanism by Lanza (2006) suggest that it cannot explain the orbital period modulation of RS CVn stars and probably all other close binary systems with a late-type secondary. As for now the Applegate hypothesis is hardly confirmed.

We have provided direct measurements (e.g. of period change rate and the O'Connell effect) for a sample of eclipsing binaries as well as interpretations of potential correlations.  The main drawback of this study is that the sample size is still too small to claim statistical significance for most of the proposed correlations.  However, we have demonstrated the use of a general-purpose photometric survey like ASAS to address these questions.  As we noted earlier, two more years of data will more than double the number of HPCR detections, so we are hopeful that the observations will be able to answer the questions we raised in the short-term.  Moreover, ASAS is set to release {\it I}-band data for all of its stars, and this color information will be useful for modeling the light-curves to constrain the stellar types, sizes, mass ratios, etc.; an exciting era of using stellar properties to infer period change mechanisms lies just ahead.

%%%%%%%%%%%%%%%%%%%%%%%%%%%%%%%%%%%%%%%%%%%%%%%%%%
\section*{Acknowledgements}
\label{sect:acknowl}
%%%%%%%%%%%%%%%%%%%%%%%%%%%%%%%%%%%%%%%%%%%%%%%%%%

We are very grateful to B. Paczy{\'n}ski, M. Kubiak, and S. Tremaine for many helpful discussions and to J. Kreiner for providing us with times of minima for SV Cen and VY Cet stars. We would also like to thank G. Pojma\'nski, K. St\c{e}pie\'n, S. Rucinski, and the anonymous referee for useful comments on the draft of this paper.  This research has made use of the SIMBAD database, operated at CDS, Strasbourg, France. This work was supported by the MNiSW grant N203 007 31/1328.  DF was supported by NASA under award No NNG04H44G to S. Tremaine.

%----------------------- Fig. 13 ---------------------------------------
%----------------------- Fig. 13 ---------------------------------------

%--------------------------------------------

\bsp
\label{lastpage}

\end{document}